\documentclass{appolb}
\usepackage{epsfig}
\begin{document}
\title{Formation of   $\Sigma \pi$  pairs in nuclear captures of K$^-$ mesons
}
\author{Raffaele Del Grande
\address{INFN Laboratori Nazionali di Frascati, Frascati, Italy\\ Universit\`a degli Studi di Roma Tor Vergata, Rome, Italy}
\\[0.5cm]{Kristian  Piscicchia}
\address{CENTRO FERMI - Museo Storico della Fisica e Centro Studi e Ricerche "Enrico Fermi", Rome, Italy\\ INFN Laboratori Nazionali di Frascati, Frascati, Italy }
 \\[0.5cm] {S{\l}awomir Wycech}
\address{National Centre  for Nuclear Studies, Warsaw,
Poland}}

\maketitle

\begin{abstract}
\abstract{The capture of K$^-$ mesons on nucleons bound in nuclei offer a chance to study the $\Sigma \pi$  pairs below the kinematic threshold of the $\bar{\mathrm{K}}$N systems. Various hyperon-pion charged combination are presently under investigation by AMADEUS. These data allow to test both isospin $0$ and $1$ amplitudes giving the possibility
to detect the structure of resonant $\Lambda(1405)$ state. Contrasted against similar electro-production data, it is possible to detect changes of  $\Lambda(1405)$ in nuclear media. Expected spectra and their uncertainties are calculated.}
\end{abstract}



\section{Introduction}
\label{intro}
The emission  of  hyperon and meson pairs $\Sigma^{\pm} \pi^{\mp} $ following the  K$^-$p capture in nuclei was   studied in nuclear emulsion  and in  bubble chambers. For example see References \cite{rif2,rif1,emuls,Fetkovich,katz,brun,riley,van-vel}. In particular the research in Ref. \cite{emuls}  concentrated
 on measurements of total P$_{\Sigma \pi} $  momenta  and invariant masses M$_{\Sigma \pi}$.  Such experiments allow to test  invariant
 mass of the K$^-$p  pair in  the sub-threshold region. One purpose of the research is to learn the structure of $\Lambda(1405) $ resonance
 located below the $\bar{\mathrm{K}}$N threshold.  Properties of the latter  may be detected with a simultaneous measurement of: M$_{\Sigma \pi}$, P$_{\Sigma\pi}$ and the ratio of two formation rates
  $\sigma(\Sigma^{+}, \pi^{-})/\sigma(\Sigma^{-}, \pi^{+})= R_{\pm}( \mathrm{M}_{\Sigma \pi})$.  This  ratio depends strongly on the invariant
  mass shapes and reflects an  interference  of the resonant isospin $0$  amplitude with a background isospin $1$.
Recent experiments  by FINUDA collaboration  allowed  more  precise measurements of
  both the  meson and the  hyperon momenta in a series of light nuclei \cite{FINU}. Unfortunately, the quantity which is of easiest use for theoretical
  analysis, the invariant mass distribution,  has not been disclosed.
AMADEUS is investigating the reactions
  \begin{equation}
\label{g0}
\mathrm{K}^- ~^{12}\mathrm{C} ~ \rightarrow  (\Sigma^{+}   ~\pi^{-}) /   (\Sigma^{0}   ~\pi^{0}) ~^{11}\mathrm{B} \ ,
\end{equation}
see References \cite{AMADEUS3,AMADEUS1,AMADEUS2}.
The experimental investigation of negative kaons absorption on ${}^{12}$C (and similar studies on $^4$He) make an extension
of the former results, offering better precision and higher statistics.
The nuclear absorptions described in equation (\ref{g0}) are due to basic transitions on  protons and are described by the following combinations of two
isospin  $ I=0,1$  transition amplitudes $T_0 $ and $T_1 $:
\begin{equation}
\label{g1}
T(\mathrm{K}^- \mathrm{p}  \rightarrow  \Sigma^{\pm} \pi^{\mp}) = \frac{1}{\sqrt{6}} T_0 \pm \frac{1}{2}T_1;  ~~T(\mathrm{K}^- \mathrm{p}  \rightarrow  \Sigma^{0} \pi^{0})= -\frac{1}{\sqrt{6}} T_0 \ .
\end{equation}
The invariant mass distribution has, in the leading approximation, the simple structure
\begin{equation}
\label{g2}
P^\mathrm{p}( \mathrm{M}_{\Sigma \pi})~\mathrm{d}\rho = |T(\mathrm{M}_{\Sigma \pi}) |^2 ~ |F^\mathrm{p}( \mathrm{P}_{\Sigma \pi})|^2~ \mathrm{d}\rho \ ,
\end{equation}
where \textquotedblleft all\textquotedblright ~ the nuclear physics is contained in a form-factor $F^\mathrm{p}$, determined by the initial state of nucleon and meson and
by the final state  interactions of the hyperon. One does not determine the absolute normalization.  The phase space element $\mathrm{d}\rho$
 makes a fairly trivial factor in the atomic capture, but becomes a bit more involved for the  in-flight captures.

 Very useful could also be a parallel experimental study of the K$^-$ capture on neutrons because the $T_1$ amplitude can be directly obtained:
\begin{equation}
\label{g3}
T(\mathrm{K}^- \mathrm{n}  \rightarrow  \Sigma^{-} \pi^{0}) = \frac{1}{\sqrt{2}} T_1 ;  ~~ T(\mathrm{K}^- \mathrm{n}  \rightarrow  \Sigma^{0} \pi^{-}) = - \frac{1}{\sqrt{2}} T_1 \ .
\end{equation}
This could be done in the context of AMADEUS. The related invariant mass distribution,
can be written, similarly to (\ref{g2}), as
\begin{equation}
\label{g21}
P^\mathrm{n}( \mathrm{M}_{\Sigma \pi})~\mathrm{d}\rho = |T(\mathrm{M}_{\Sigma \pi}) |^2 ~ |F^\mathrm{n}( \mathrm{P}_{\Sigma \pi})|^2~ \mathrm{d}\rho \ ,
\end{equation}
where $F^\mathrm{n}$ is now the form-factor for the K$^-$n interactions. Assuming that $|F^\mathrm{n}|  \simeq |F^\mathrm{p}|$, the nuclear physics can be disentangled and the ratio
\begin{equation}
R(\mathrm{p/n}) = \frac{P^\mathrm{p}( \mathrm{M}_{\Sigma \pi})}{P^\mathrm{n}( \mathrm{M}_{\Sigma \pi})} \ ,
\end{equation}
allows to directly study the ratio between $T_0$ and $T_1$. The main complication in such analysis comes from the $I=1$ $\Sigma(1385)$ resonance formation in $P$ wave K$^-$n interaction. Indeed, while in $S$ wave the $T_1$ can be considered approximately constant, the  $P$ wave K$^-$n interaction is affected by the resonance formation. However, the $\Sigma(1385)$ is fortunately very weakly coupled to the $\Sigma \pi$ decay channel. This analysis is also complicated by the final state absorption of the hyperon, which will be discussed in section 3.

The ratio $R_\pm (\mathrm{M}_{\Sigma \pi}) = P^\mathrm{p}(\mathrm{M}_{\Sigma^+ \pi^-})/ P^\mathrm{p}(\mathrm{M}_{\Sigma^- \pi^+})$ was studied in the thesis by Keane \cite{Keane} and was analysed in Ref. \cite{Staro} where also the data are reported. These, not very precise, data give a ratio which indicates an anomaly
 35 MeV below the  K$^-$p threshold which   was  discussed  in terms of  Dalitz \cite{Dalit} suggestion  of  a sizable three quark component in $\Lambda(1405)$.
   Recent electro-production experiments on proton indicate $R_\pm (\mathrm{M}_{\Sigma \pi})$  to be a fairly smooth function of the energy \cite{jeff}. Thus the anomaly in question
   is apparently related to the presence of the nucleus. A second anomaly can be found in the data published by FINUDA \cite{FINU}, where a strong enhancement of events close to the $\Sigma^+$ formation threshold, that is for low $\Sigma^+$ energies, can be observed for K$^-$ captures at-rest on ${}^{6}$Li target. A Monte Carlo interpretation in terms of energy loss of the $\Sigma^+$ in the target seems to miss an accurate description of the measured P$_{\Sigma^+}$ momentum spectra. AMADEUS also reported, in Ref. \cite{AMADEUS2}, a low momentum peak structure in P$_{\Sigma^+}$ momentum distribution, in a sample of $\Sigma^+ \pi^-$ pairs produced in K$^- ~ {}^{12}$C absorptions. The low energy ${\Sigma^+}$ events amount to some percent of the total sample. The solid target is much thiner in this case, so again energy loss seem not the only satisfying explanation. Moreover the low momentum structure is not observed in \cite{AMADEUS2} when the K$^-$ is absorbed on a solid ${}^{9}$Be target. These findings are interpreted in Ref. \cite{WKthis} as formation of a Gamov state in the $\Sigma^+$-residual nucleus  system. It would be of interest to learn if a relation among the two anomalies hold. In this note we present the \textquotedblleft gross structure\textquotedblright ~ of the spectra.
In particular  we present technical description of the  distributions $P^\mathrm{p}(\mathrm{M}_{\Sigma \pi})$ while the  hyperon momentum distribution $P(\mathrm{P}_{\Sigma} )$ is presented in a parallel work \cite{WKthis}.

\section{Emission probabilities}

The $\bar{\mathrm{K}}$N forces are known to be very short, we then use  the transition operator of zero range. This assumption allows to write the capture amplitude $A$ as follow:
\begin{equation}
\label{g4}
A =  \int \mathrm{d}\textbf{r} ~\Phi^*_\Sigma (r)\Phi^*_\pi(r) T(\mathrm{K}^- \mathrm{p} \rightarrow \Sigma\pi) \Phi_\mathrm{p}(r) ~ \Phi_\mathrm{K}(r) \sim T(\mathrm{M}_{\Sigma\pi}) ~ F^\mathrm{p} (\mathrm{P}_{\Sigma\pi}) \ ,
\end{equation}
where $F^\mathrm{p}$ is the form-factor introduced in (\ref{g2}) and defined as
\begin{equation}
\label{g41}
F^\mathrm{p} (\mathrm{P}_{\Sigma\pi}) =  \int \mathrm{d}\textbf{r} ~\Phi^*_\Sigma (r)\Phi^*_\pi(r) ~ \Phi_\mathrm{p}(r) ~ \Phi_\mathrm{K}(r) \ .
\end{equation}
The definition (\ref{g41}) requires the knowledge of the initial wave function
of the proton $\Phi_\mathrm{p}(r)$ (taken from Ref. \cite{neff}) and Kaon $\Phi_\mathrm{K}$ (calculated with K-nucleus optical potential). Due to peripherality of the absorption, it occurs essentially on the P-wave nucleons. As the absolute rate is not measured the overlap of initial and final nuclei is not relevant in the determination of the spectra.
The wave function of the Kaon depends on the atomic quantum numbers $n$ and $l$ of the orbital from which the K$^-$ is captured.
For  captures in flight $\Phi_\mathrm{K}$  is close to a plane wave, and the  initial state is known.
The X-ray transitions in Carbon terminate at $l=1$ state but the $l = 2$ is apparently  the dominant angular momentum at the capture. The absolute rate of radial  $l=3 \rightarrow l=2 $ transition is  0.36(6) while the rate of subsequent radial transition is only 0.028(8)
\cite{poth}. 
The distribution in terms of  main quantum numbers is not known
but it  is not relevant as  the absolute capture rates are not measured. With the nuclear oscillator model of parameter $\alpha$  and pure  Coulomb atoms
 one   obtains:
\begin{equation}
\label{g6}
<| F^\mathrm{p} (\mathrm{P}_{\Sigma \pi}) |^2 > ~ \sim   ~ \frac{ \mathrm{P}_{\Sigma \pi}^2}{\alpha^2}
 \ e^{-\frac{\mathrm{P}_{\Sigma \pi}^2}{\alpha^2}} \
 \left[8 \left( \frac{5}{2}-\frac{ \mathrm{P}_{\Sigma \pi}^2}{2\alpha^2} \right)^2 +\frac{3 \mathrm{P}_{\Sigma \pi}^4}{\alpha^4} \right] \ ,
\end{equation}
where the form-factor is averaged over the atomic and nuclear magnetic orientation and summed over protons.
The phase space d$\rho$ in eq. (\ref{g2}) is given by
$$ \mathrm{d}\rho =  \rho \ \mathrm{d M}_{\Sigma\pi}=  \mathrm{P}_{\Sigma \pi} \sqrt{ \mathrm{E}_0-\mathrm{M}_\Sigma -\mathrm{M}_\pi + \frac{\mathrm{P}_{\Sigma \pi}^2}{2(\mathrm{M}_\Sigma + \mathrm{M}_\pi)}}~\mathrm{d M}_{\Sigma\pi}  \ ,$$
where  E$_0$ = M$_\mathrm{p}$ + M$_\mathrm{K} -$ E$_\mathrm{p}^\mathrm{binding}$, E$_\mathrm{p}^\mathrm{binding}$ is the binding energy of the absorbing proton (and Kaon),  for relativistic corrections we refer to \cite{KWC}.
The final probability distribution function is
\begin{equation}
\label{g7}
P(\mathrm{M}_{\Sigma\pi}) \mathrm{d M}_{\Sigma\pi} =  |T(\mathrm{M}_{\Sigma\pi})|^2 <| F^\mathrm{p} (\mathrm{P}_{\Sigma\pi}) |^2 > ~ \rho~\mathrm{d M}_{\Sigma\pi}  \ .
\end{equation}
It turns out that valence protons contribute $90\%$ of the rate and this simplifies  the relation of the invariant mass to the  momentum $( \mathrm{M}_{\Sigma\pi})^2 =  \mathrm{E}_0^2 - \mathrm{P}_{\Sigma\pi}^2$ as  the $\mathrm{E}_\mathrm{p}^\mathrm{binding}$  differs strongly in   $2p$ where $\mathrm{E}^\mathrm{binding} = 16$ MeV and $1s$  nucleon orbital where $\mathrm{E}^\mathrm{separation} \simeq 30-50$ MeV. In the atomic capture, the kinematic limit on the achievable invariant mass is 1416 MeV. For in flight captures, the upper limit is pushed up of $\sim$ 14 MeV, by the kinetic energy of the Kaon, for Kaon momenta of $\sim$ 120 MeV.

\begin{figure}[htb]
\centerline{\includegraphics[height= 5cm,width=7cm]{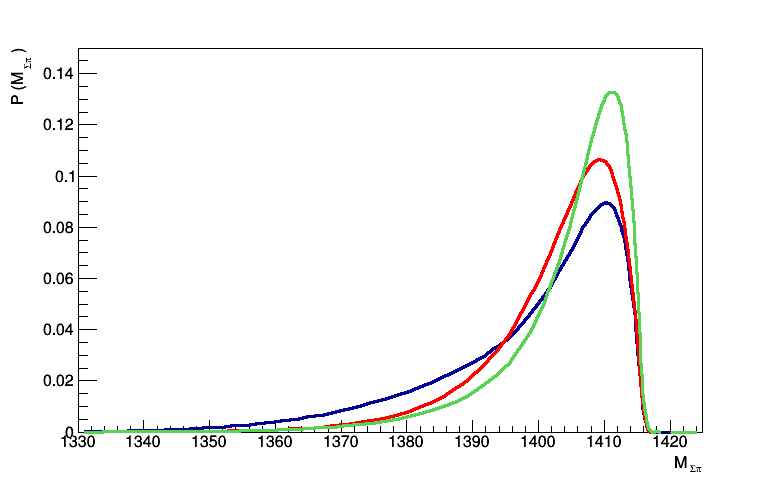}}
\caption{Invariant mass  distribution of  $\Sigma^+ \pi^-$  pairs following K$^-$  capture from atomic $l=2$ state in Carbon. Final state interactions have not been calculated.
The curves test the dependence of the spectrum on the position of a 40 MeV wide resonance centred at E$_r = 1405$ MeV ($red ~ curve$), E$_r = 1420$ MeV ($green ~ curve$). The non-resonant shape ($blue$) is obtained using $| T(\mathrm{M}_{\Sigma\pi})|^2  = 1$. The area of the three curves is normalized to unity.
}
\label{f2}
\end{figure}

Figure 1 displays profiles of the $\Sigma^+ \pi^-$ invariant mass spectrum including effects of resonant K$^-$p interactions. Final state interactions have not been calculated. The shapes are obtained from the probability distribution function in eq. (\ref{g7}), using $|T(\mathrm{M}_{\Sigma\pi})|^2  = 1$ in non-resonant reaction while a Breit-Wigner shape is used for $|T(\mathrm{M}_{\Sigma\pi})|^2$ in the resonant reaction. The difficulty of extraction of the resonance is due essentially to the sharp cut at 1416 MeV due to phases space limitations. However, the profiles 
are distinctive enough to allow checks of $T(\mathrm{M}_{\Sigma\pi})$.

\section{Higher order effects}
Several  corrections should be kept under control when experimental results are analyzed:
\begin{itemize}
\item[1)] \emph{\underline{Initial state meson interactions}}.
In atomic capture, the correction due to the initial interaction of the meson is easy to introduce since the optical potential is  known from X-ray data \cite{elif}.
Moreover, it is the same for all $\Sigma \pi$ pairs and drops out when studying the ratios of emission rates.
For captures in flight the analysis is more difficult because informations from only one, old, scattering experiment on $^4$He target \cite{mazur}, are available. In addition, such experiment does not agree with the atomic data. This may be due to the rapid change of the resonant $\bar{\mathrm{K}}$N amplitude, or to the poor energy resolution and then to the low quality of the data. 
Such alternative might be solved by the in flight experiment on C and in better way on He targets.

\item[2)]  \emph{\underline{The final hyperon absorptive interactions}}. This correction is more difficult to implement as it depends on the final charged channel. In particular,
it is known from emulsion works that absorption of final $\Sigma^+$ differs from the absorption of final $\Sigma^-$  \cite{emuls,Staro}.
It is due to the Coulomb and to differences in poorly known optical potentials for these  hyperons, both affect the low energy
part of the hyperon momentum spectra. Part of the difference is also related to the effect of the Gamov state, which is discussed in Ref. \cite{WKthis}.
The description of such differences is rather involved and not very reliable \cite{Staro}.  Help from experiment is needed as the emulsion work
provides total absorption rates  and no related energy dependence. The best way for this experiment is to extract the difference from the hyperon momentum
spectra.

\item[3)]  \emph{\underline{Subtler effects related to the nuclear structure}}. These effects are given by the fact that, in reaction (\ref{g0}), the final
$^{11}$B is a third body spectator. So far we included the related energy recoil. Other  changes would be to replace the  coordinate $\textbf{r}$
by a pair of  Jacobi coordinates as done in the parallel calculation \cite{WKthis}. That part is simple in an oscillator model of the nucleus \cite{neff}.
\end{itemize}

\end{document}